\documentclass[aps,prb,twocolumn,superscriptaddress]{revtex4-2}

\usepackage{amsmath, amssymb,  graphicx}%, subcaption}

\usepackage[colorlinks=true ,urlcolor=blue,urlbordercolor={0 1 1}]{hyperref}

\usepackage{color}
\usepackage{xcolor}
\usepackage{braket}

\usepackage[utf8]{inputenc}
\usepackage{bm}
\usepackage{verbatim}
\usepackage{graphicx}
\usepackage{amsmath}
\usepackage{color}
\usepackage{float}
\usepackage{bbm}
\usepackage{amssymb}
\usepackage{slashed}
\usepackage{wasysym,bm,bbm,dsfont}
\usepackage{soul}

\begin{document}

\title{New classes of quantum anomalous Hall crystals in multilayer graphene}
\author{Boran Zhou}
\author{Ya-Hui Zhang}

\affiliation{Department of Physics and Astronomy, Johns Hopkins University, Baltimore, Maryland 21218, USA}

\date{\today}

\begin{abstract}
The recent experimental observation of  quantum anomalous Hall (QAH) effects in the rhombohedrally stacked pentalayer graphene has motivated theoretical discussions on the possibility of quantum anomalous Hall crystal (QAHC), a topological version of Wigner crystal. Conventional topological Wigner crystals typically have one electron per unit cell. In this work we propose new types of topological Wigner crystals labeled as QAHC-$z$, with $z$ electrons per unit cell.  In the pentalayer graphene system, we find parameter regimes where QAHC-2 and QAHC-3 have lower energy than the conventional QAHC-1 at total filling $\nu=1$ per moir\'e unit cell. These states all have total Chern number $C_\mathrm{tot}=1$ and are consistent with the QAH effect observed in the experiments. The larger period QAHC states have lower kinetic energy due to the unique Mexican-hat dispersion of the pentalayer graphene, which can compensate for the loss in the interaction energy. Unlike QAHC-1, QAHC-2 and QAHC-3 break the moir\'e translation symmetry and are sharply distinct from a moir\'e band insulator. We also briefly discuss the competition between integer QAH and fractional QAH states at filling $\nu=2/3$. Moreover, we find that a stronger moir\'e potential can significantly change the phase diagram and even favors a QAHC-1 ansatz with $C=2$ Chern band.
\end{abstract}

\maketitle

\textbf{Introduction}
Fractional quantum anomalous Hall~(FQAH) effects, defined as fractional quantum Hall (FQH) effects \cite{stormer1999fractional,jain1989composite} at zero magnetic field, have been theoretically proposed to be  possible in a narrow Chern band \cite{sun2011nearly,sheng2011fractional,neupert2011fractional,regnault2011fractional,tang2011high,wang2011fractional,bergholtz2013topological,parameswaran2013fractional,PhysRevB.84.165107}.  In recent years, various moir\'e systems have been proposed as promising candidates \cite{zhang2019nearly,wu2019topological,ledwith2020fractional,repellin2020chern,abouelkomsan2020particle,wilhelm2021interplay,li2021spontaneous,crepel2023fci,devakul2023magic,gao2023untwisting,ghorashi2023topological}. Integer quantum anomalous Hall (QAH) effects have  been reported in twisted bilayer graphene (TBG) aligned with hexagon boron nitride (hBN) \cite{sharpe2019emergent,serlin2020intrinsic}, ABC-trilayer graphene/hBN moir\'e system \cite{chen2020tunable}, transition metal dichalcogenide (TMD) moir\'e heterobilayers \cite{li2021quantum} and homobilayers \cite{doi:10.1126/science.adi4728}. Later,  FQAH was finally realized in twisted MoTe$_2$ homobilayers \cite{cai2023signatures,zeng2023thermodynamic,park2023observation,PhysRevX.13.031037} and rhombohedrally stacked pentalayer graphene \cite{lu2023fractional,waters2024interplay}.  In twisted MoTe$_2$, the first valence band has a nonzero Chern number $C=1$ \cite{wu2019topological,yu2020giant,devakul2021magic}, resembling a Landau level, 
 which makes FQAH states at fractional fillings quite natural.  In contrast, in the pentalayer graphene system, even the integer QAH effect at $\nu=1$ is a surprise because the moir\'e potential is too weak to open a band gap.

 Theoretical progress has been made to understand the integer filling $\nu=1$ for pentalayer graphene \cite{dong2023theory,zhou2023fractional,dong2023anomalous,herzog2024moire,PhysRevB.110.075109,kwan2023moir,yu2024moir,PhysRevB.109.L241115,huang2024fractional}. Although there is no band gap at the single particle level because electrons are pushed to layers far away from the aligned hBN and feel only a weak moir\'e potential, Hartree Fock calculations \cite{dong2023theory,zhou2023fractional,dong2023anomalous,PhysRevB.110.075109,kwan2023moir} show that interactions can spontaneously generate a large crystal potential at the mean field level, providing a narrow Chern band with $C=1$. In the limit of a vanishing moir\'e potential, the phase is a topological version of Wigner crystal, dubbed as quantum anomalous Hall crystal (QAHC) \cite{zhou2023fractional,dong2023anomalous}, while in some reference QAHC is referred to as a topological charge density wave~\cite{PhysRevLett.133.066601,PhysRevB.109.115116}. Similar physics is predicted to appear in rhombohedrally stacked $n_\mathrm{layer}$-layer graphene with $n_\mathrm{layer}=4,6,7$, later verified by experiments \cite{lu2024extended,choi2024electric,xie2024even,ding2024electrical}. The new concept of QAHC has attracted lots of attentions~\cite{PhysRevLett.132.236601,soejima2024anomalous,dong2024stability,tan2024parent,tan2024wavefunction,patri2024extended,PhysRevB.110.155148,wei2024edge}.  However, there are debates on whether the QAH insulator is really different from a moir\'e band insulator and the role of the moir\'e potetial~\cite{dong2023anomalousjournalclub}. In this paper we ask the following question: can we find a QAHC state which also breaks the moir\'e translation symmetry, thereby sharply distinguished from a moir\'e band insulator? If an integer QAH effect is realized at fractional filling, it is guaranteed to break translation symmetry, as observed in twisted bilayer-trilayer graphene \cite{su2024generalized}. Here we present surprising results showing that even the integer QAH insulator at $\nu=1$ may break moir\'e translation symmetry.

Given that the moir\'e potential is a perturbation, we can first consider the moir\'eless limit.  We extend the  quantum anomalous Hall crystal concept and use QAHC-$z$ to label a crystal with a unit cell size $z$ times larger than the usual QAHC or Wigner crystal. Specifically, the crystal period is given by $a_\mathrm{crystal}=(\frac{4z^2}{3})^{1/4}n^{-1/2}$, where $n$ is the electron density. Previous Hartree Fock calculations were restricted to QAHC-1 without any enlargement of the unit cell at $\nu=1$.  In contrast, we also include the possibilities of QAHC-2 and QAHC-3 in our new Hartree Fock calculations and find that they can be favored in certain parameter space spanned by the twist angle and the displacement field $\delta_D$. These new crystals have 2 or 3 bands fully occupied, but still have a total Chern number $C_\mathrm{tot}=1$, consistent with the QAH effect with $\sigma_{xy}=e^2/h$. Unlike QAHC-1, they break the moir\'e translation symmetry and are unambiguously beyond band insulator descriptions.

We also study the competition between integer QAHCs and FQAH states at fractional filling $\nu=2/3$. Following the spirit of the variational method, we compare the energy of an FQAH state obtained by exact diagonalization (ED) and QAHCs from Hartree Fock calculations.  We find that the FQAH state is usually at a lower energy and is increasingly favored with stronger interaction strengths and a stronger moir\'e potential. Among the integer QAH states, the QAHC-1 and QHAC-2 are in close competitions, with either one being potentially favored depending on parameters. Hence we expect a rich phase diagram in rhombohedral multilayer graphene systems.

\textbf{Model}
We employ the continuum model to calculate the bare band structure, with the Hamiltonian for each valley and spin given by:
\begin{equation}\label{eqHk}
    H_K = H_0 + V_M H_M,
\end{equation}
where $H_0$ represents the Hamiltonian of rhombohedral pentalayer graphene:
\begin{equation}
        H_0 = 
    \begin{pmatrix}
    H_1 & \Gamma & \tilde{\Gamma} & \mathbf{0_{2\times 2}} & \mathbf{0_{2\times 2}} \\
    \Gamma^\dagger & H_2 & \Gamma & \tilde{\Gamma} & \mathbf{0_{2\times 2}} \\
     \tilde{\Gamma}^\dagger   & \Gamma^\dagger & H_3 & \Gamma & \tilde{\Gamma}  \\
    \mathbf{0_{2\times 2}} & \tilde{{\Gamma}}^\dagger & \Gamma^\dagger & H_4 & \Gamma \\
   \mathbf{0_{2\times 2}}  & \mathbf{0_{2\times 2}} & \tilde{{\Gamma}}^\dagger & \Gamma^\dagger & H_5
    
    \end{pmatrix},
\end{equation}
with the block matrices defined as follows:
\begin{eqnarray}
    &H_i=\begin{pmatrix}\label{eqH03}
        (\frac{i}{4}-\frac{1}{2})\delta_D +u_{A,i}& -\frac{\sqrt{3}}{2}\gamma_0  (\tilde{k}_x-\mathrm{i}\tilde{k}_y)\\
        -\frac{\sqrt{3}}{2}\gamma_0 (\tilde{k}_x+\mathrm{i}\tilde{k}_y) & (\frac{i}{4}-\frac{1}{2})\delta_D+u_{B,i}
    \end{pmatrix},&\\
    &\Gamma=\begin{pmatrix}
        -\frac{\sqrt{3}}{2}\gamma_4 (\tilde{k}_x-\mathrm{i}\tilde{k}_y) & -\frac{\sqrt{3}}{2}\gamma_3 (\tilde{k}_x+\mathrm{i}\tilde{k}_y) \\
        \gamma_1 &  -\frac{\sqrt{3}}{2}\gamma_4  (\tilde{k}_x-\mathrm{i}\tilde{k}_y)        
    \end{pmatrix},&\\
    &\tilde{\Gamma}=
    \begin{pmatrix}
        0 & \frac{1}{2}\gamma_2 \\
        0 & 0
    \end{pmatrix}. &
\end{eqnarray}
The parameters used are $(\gamma_0,\gamma_1,\gamma_2,\gamma_3,\gamma_4)=(-2600,358,-8.3,293,144)$~meV. The potential difference between the top and the bottom graphene layers is denoted by $\delta_D$. $u_{A,1}=u_{B,5}=0$, $u_{B,1}=u_{A,5}=12.2$~meV, $u_{A,i}=u_{B,i}=-16.4$~meV for other terms. We rotate the graphene as $\tilde{k}_x+\mathrm{i}\tilde{k}_y=e^{\mathrm{i}\theta_3}(k_x+\mathrm{i}k_y)$, where $\theta_3=\arctan\frac{\theta}{\delta_a}$ with $\theta$ representing the twist angle between the graphene and the hBN. $\delta_a=\frac{a_\mathrm{hBN}-a_\mathrm{G}}{a_\mathrm{hBN}}=0.017$, where $a_\mathrm{hBN}$ and $a_\mathrm{G}$ are the lattice constants of graphene and hBN respectively.

The moir\'e potential, representing the tunneling between the first graphene layer aligned with the hBN, is given by:
\begin{equation}
    H_M(\mathbf{G}_j) = 
    \begin{pmatrix}
        C_0 e^{\mathrm{i}\phi_0}+C_z e^{\mathrm{i}\phi_z} & C_{AB}e^{\mathrm{i}(\frac{(3-j)\pi}{3}-\phi_{AB})}\\
        C_{AB}e^{\mathrm{i}(\frac{(1+j)\pi}{3}-\phi_{AB})} & C_0e^{\mathrm{i}\phi_0}-C_ze^{\mathrm{i}\phi_z}
    \end{pmatrix},
\end{equation}
where the momentum difference given by $\mathbf{G}_j=\frac{4\pi}{\sqrt{3}L_M}(\cos (\frac{j\pi}{3}-\frac{5\pi}{6}),\sin (\frac{j\pi}{3}-\frac{5\pi}{6}))^T$ for $j=1,3,5$. For $j=2,4,6$, the tunneling is obtained by taking the Hermitian conjugate. The parameters are determined from DFT calculations~\cite{PhysRevB.108.155406} as $C_0=-10.13~\mathrm{meV}$, $\phi_0=86.53^\circ$, $C_z=-9.01~\mathrm{meV}$, $\phi_z=8.43^\circ$, $C_{AB}=11.34~\mathrm{meV}$, $\phi_{AB}=19.60^\circ$. Here we introduce an artificial parameter $V_M$ in Eq.~
\ref{eqHk}. We project the moiré potential onto the conduction bands to compare the energies for different $V_M$ values (see the Supplemental Material). 

The interaction is given by the Coulomb potential:
\begin{equation}
    H_V=\frac{1}{2A}\sum_{l,l^\prime}\sum_\mathbf{q}V_{ll^\prime}(\mathbf{q}):\rho_l(\mathbf{q})\rho_{l^\prime}(-\mathbf{q}):,
\end{equation}
in which $A$ is the area of the system, $\rho_l(\mathbf{q})$ is the density in the layer $l$. $V_{ll^\prime}(\mathbf{q})=\frac{e^2\tanh q\lambda}{2\epsilon_0\epsilon q}e^{-q|l-l^\prime|d_\mathrm{layer}}$, $\lambda=30$~nm is the screening length and $d_\mathrm{layer}=0.34$~nm is the distance between adjacent layers. In all of the calculations, we use $\epsilon=10$ unless specified.

\textbf{Integer QAHC at $\nu=1$}
For the integer QAH insulator at $\nu=1$ observed in recent experiments~\cite{lu2023fractional,waters2024interplay,lu2024extended}, we focus on the conduction bands. Due to the large and negative displacement field $\delta_D$, the conduction electrons mainly stay in the bottom layer, away from the aligned hBN on the top. As a consequence, the moir\'e potential projected onto the conduction band is less than $0.05$~meV~\cite{zhou2023fractional}. The previous theories suggest that the insulator at $\nu=1$ is a QAHC state~\cite{dong2023theory,zhou2023fractional,dong2023anomalous} that is pinned by this small moir\'e potential. However, previous studies have been limited to crystals whose period is the same as the moir\'e period. In principle, the topological crystal state can have different crystal periods.  We generalize the notion of QAHC to QAHC-$z$, where $z\in\mathbb{Z}$, and the crystal period is labeled as $a_\mathrm{crystal}=\sqrt{z/\nu}a_M$. The corresponding reciprocal lattice vector is represented as $\mathbf{G}^{(z)}_i$. For values of $z/\nu=m_1^2+m_2^2+m_1m_2$, where $m_1,m_2\in\mathbb{Z}$, the moir\'e potential is commensurate with the crystal period, allowing it to pin the crystal. For other values of $z$, the ground state energy remains unaffected by the moir\'e potential when treated as a first order perturbation. 

\textbf{Phase diagram of QAH crystals}
We assume fully spin and valley polarization \cite{zhang2019nearly,repellin2020ferromagnetism} and perform HF calculations at $\nu=1$ for different QAHC-$z$ ansatzes. For each QAHC-$z$, the calculation is conducted in a smaller Brillouin zone (BZ) scaled by $1/\sqrt{z}$ relative to the moir\'e Brillouin zone (MBZ), with $z$ electrons per unit cell. We compare the energies of different crystals with $z=1,2,3$ and select the one with the lowest energy. The resulting phase diagram is shown in Fig.~\ref{fig:phase_diagram_finite_size}(a). The anomalous Hall conductivity is determined by  the total Chern number $C_\mathrm{tot}$ for the fully filled bands, i.e., first $z$ HF conduction bands for QAHC-$z$. $C_\mathrm{tot}$ is $0$ or $1$ in the whole phase diagram, consistent with previous studies restricted to the $z=1$ ansatz \cite{dong2023theory,zhou2023fractional,dong2023anomalous}. However, we now find two additional phases at $V_M=1$: QAHC-$2$ and QAHC-$3$ with also $C_\mathrm{tot}=1$. Moreover, we manually increase the moir\'e potential by setting $V_M$ to be 2 and 5 in Fig.~\ref{fig:phase_diagram_finite_size}(b)(c) to study the effect of a larger moir\'e potential. We note that QAHC-1 is favored by larger $V_M$ as expected.  For $V_M=5$, there is also a $C_\mathrm{tot}=2$ QAHC-$1$ state in some region of $(a_M,\delta_D)$, indicating the possibility of realizing higher Chern number in the pentalayer rhombohedral graphene system by enhancing the moir\'e potential. We also include results for a different parameter with $|\gamma_0|=3100$ meV in the Supplemental Material. There we find QAHC-2 occupies an even larger region for $V_M=1$. Also for $V_M=2$, we can already find QAHC-1 with $C=2$. Although it is still not clear which parameter is more realistic, we can conclude that the QAHC-2 phase is possible and the $C=2$ QAH insulator can be potentially stabilized with sufficiently larger $V_M$. 

To distinguish different QAHC-$z$ states, one can measure the density distributions of conduction electrons, which is calculated as follows:

 \begin{figure}[tbp]
    \includegraphics[width=0.9\linewidth]{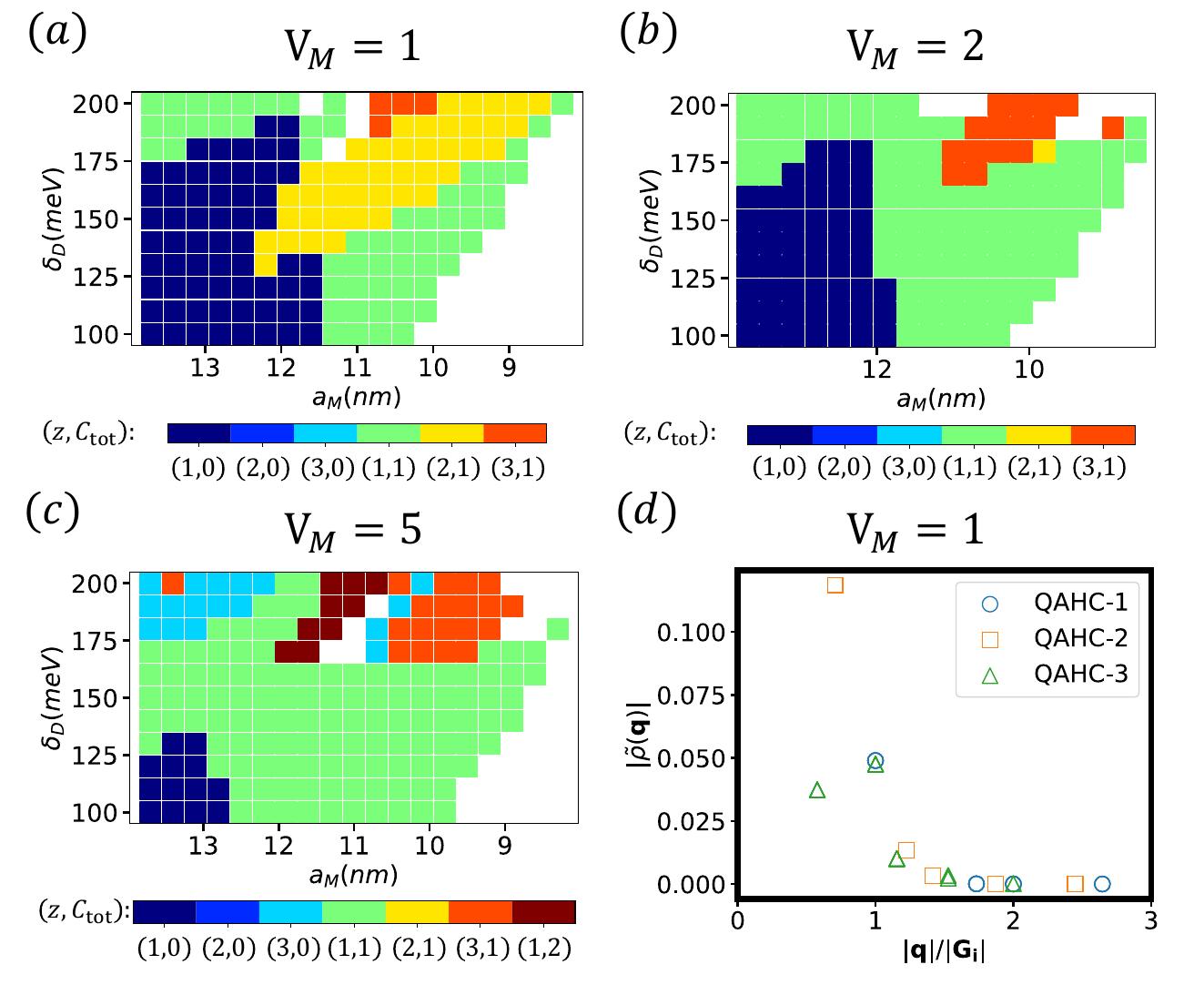}
    \caption{HF calculation results at $\epsilon=10$. (a) The phase diagram of QAHC as a function of displacement field $D$ and moir\'e period $a_M$ and $V_M=1$. The label $(z,C_\mathrm{tot})$ represents the state is QAHC-$z$ with total Chern number $C_\mathrm{tot}$. We keep $N_{b,c}$ conduction bands and $N_{b,v}$ valence bands in the calculation. For QAHC-$z$ with $z=1,2,3$, we set $N_{b,c}$ to 7, 7 and 9 respectively. $N_{b,v}$ is taken as $0$ for all values of $z$, since the valence bands are separated by the large displacement field. The calculations are performed using an $N_k\times N_k$ grid in the crystal BZ. For $z=1,2,3$, we set $N_k$ to 12, 12 and 9 respectively. (b)(c) Similar to (a), but with $V_M=2,5$ instead. (d) For $\theta=0.89^\circ$ ($a_M=10.7$~nm) and $V_M=1$, displays of the function $|\tilde{\rho}(\mathbf{q})|$ of QAHC-$1$ ($\delta_D=120$~meV), QAHC-$2$ ($\delta_D=160$~meV) and QAHC-$3$ ($\delta_D=210$~meV) respectively.  }
    \label{fig:phase_diagram_finite_size}
\end{figure}
% (c) The energy difference per moir\'e unit cell $E_\mathrm{QAHC1}-E_\mathrm{QAHC2}$ varies with $1/N_{b,c}$. The blank space corresponds to the metal phase. The energy difference converges to a positive value as $N_{b,c}\rightarrow\infty$. (d) The energy per moir\'e unit cell for QHAC-$z$ with $z=1,2,3$ is plotted as a function of $1/N_k$. \yhz{Obviously the x axis range should start from 0.}

\begin{eqnarray}
    \rho(\mathbf{r})&=&\sum_\mathbf{q}\tilde{\rho}(\mathbf{q})e^{-\mathrm{i}\mathbf{q\cdot r}},\\
    \tilde{\rho}(\mathbf{q})&=&\frac{1}{nN_k^2}\sum_{m_1,m_2,\mathbf{k}}\langle u_{\mathrm{HF},m_1}(\mathbf{k+q})|u_{\mathrm{HF},m_2}(\mathbf{k})\rangle,
\end{eqnarray}
where $k$ is defined in the crystal BZ, which is discretized into $N_k\times N_k$ points. Here, $|u_{\mathrm{HF},m_1}(\mathbf{k})\rangle$ represents the periodic part of the $m_1^\mathrm{th}$ conduction HF Bloch wavefunction, and $m_1,m_2$ range over the values $0,1,...,N_{b,c}-1$. $\mathbf{q}=n_1\mathbf{G_1}+n_2\mathbf{G_2}$ corresponds to any integer linear combination of the crystal reciprocal lattice vectors. In Fig.~\ref{fig:phase_diagram_finite_size}(d), we compare $\tilde{\rho}(\mathbf{q})$ for different QAHC-$z$ states, showing that $|\tilde{\rho}(\mathbf{G}_i^{(z)})|$ values are on the same scale. Future experiments may identify the specific state of the system under given parameters by measuring $\tilde{\rho}(\mathbf{q})$. Moreover, we present the real space density profile $\rho(\mathbf{r})$ for QAHC-1, QAHC-2, QAHC-3 in Fig.~\ref{fig:density_profile}(a)-(c), respectively. In addition, we plot the density profile for the $C=2$ state at $V_M=5$  in Fig.~\ref{fig:density_profile}(d). One can clearly see the different crystal periods and shapes of these phases at the same filling $\nu=1$.  The corresponding band structures are shown in Fig.~\ref{fig:dispersion}.

% The huge difference among QAHC-$z$ density distributions provide the new candidates of phases in this system.
 \begin{figure}[tbp]
 \includegraphics[width=0.9\linewidth]{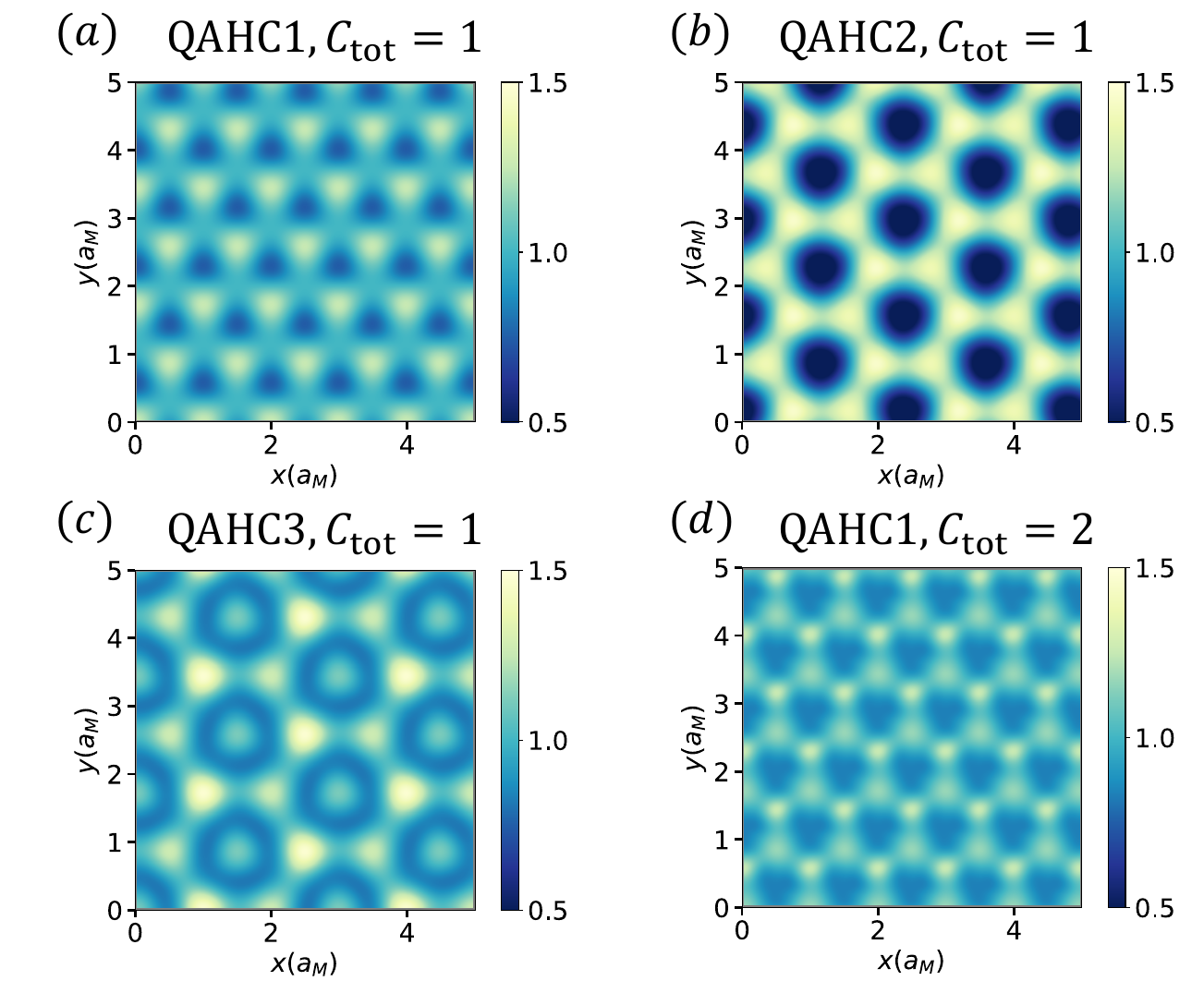}
    \caption{In (a)-(c), $\theta=0.89^\circ$ ($a_M=10.7$~nm). The density profiles are shown for (a) ($\delta_D=120$~meV, QAHC-$1$), (b) ($\delta_D=160$~meV, QAHC-$2$) and (c) ($\delta_D=210$~meV, QAHC-$3$). In (d), we manually set $V_M=5$ and plot the density distribution for $\theta=0.78^\circ$ ($a_M=11.3$~nm), $\delta_D=190$~meV.  }
    \label{fig:density_profile}
\end{figure}

 \begin{figure}[tbp]
 \includegraphics[width=0.9\linewidth]{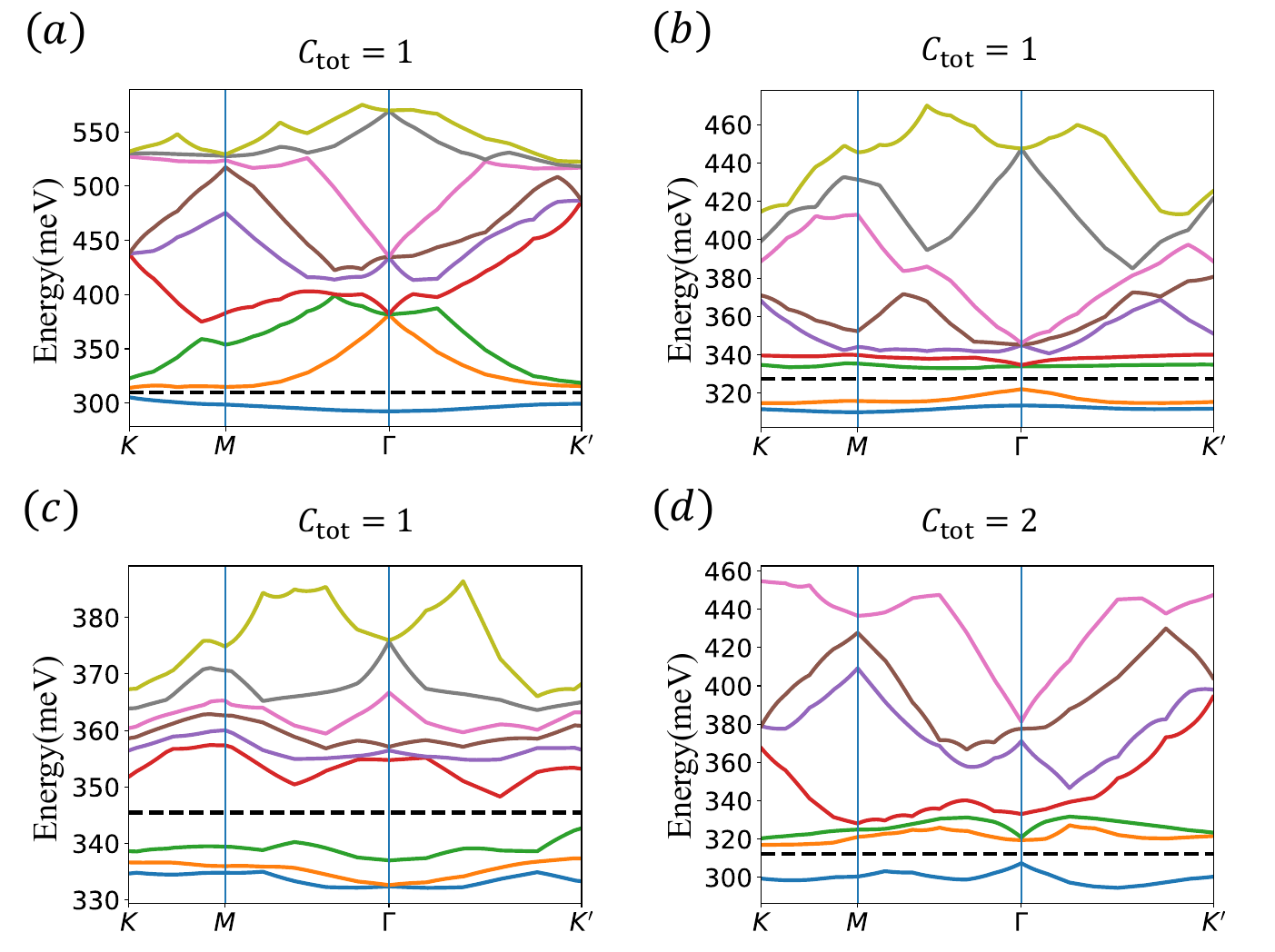}
    \caption{The parameters used in (a)-(d) is the same as in Fig.~\ref{fig:density_profile} (a)-(d). The black dashed line represents the Fermi energy.}
    \label{fig:dispersion}
\end{figure}

% From the theoretical side, we choose a specific parameter with $\delta_D=170$~meV, $\theta=0.77^\circ$ in the HF calculation to check the convergence of calculation. As shown in Fig.~\ref{fig:phase_diagram_finite_size}(c)(d), The energy difference converges as $N_{b,c}\rightarrow\infty$ and $N_k\rightarrow\infty$, respectively. \textbf{The result indicates that the error due to finite size effects is less than $5\%$ of the energy difference, confirming the reliability of the calculation.}

\begin{figure}[tbp]
 
    \includegraphics[width=0.9\linewidth]{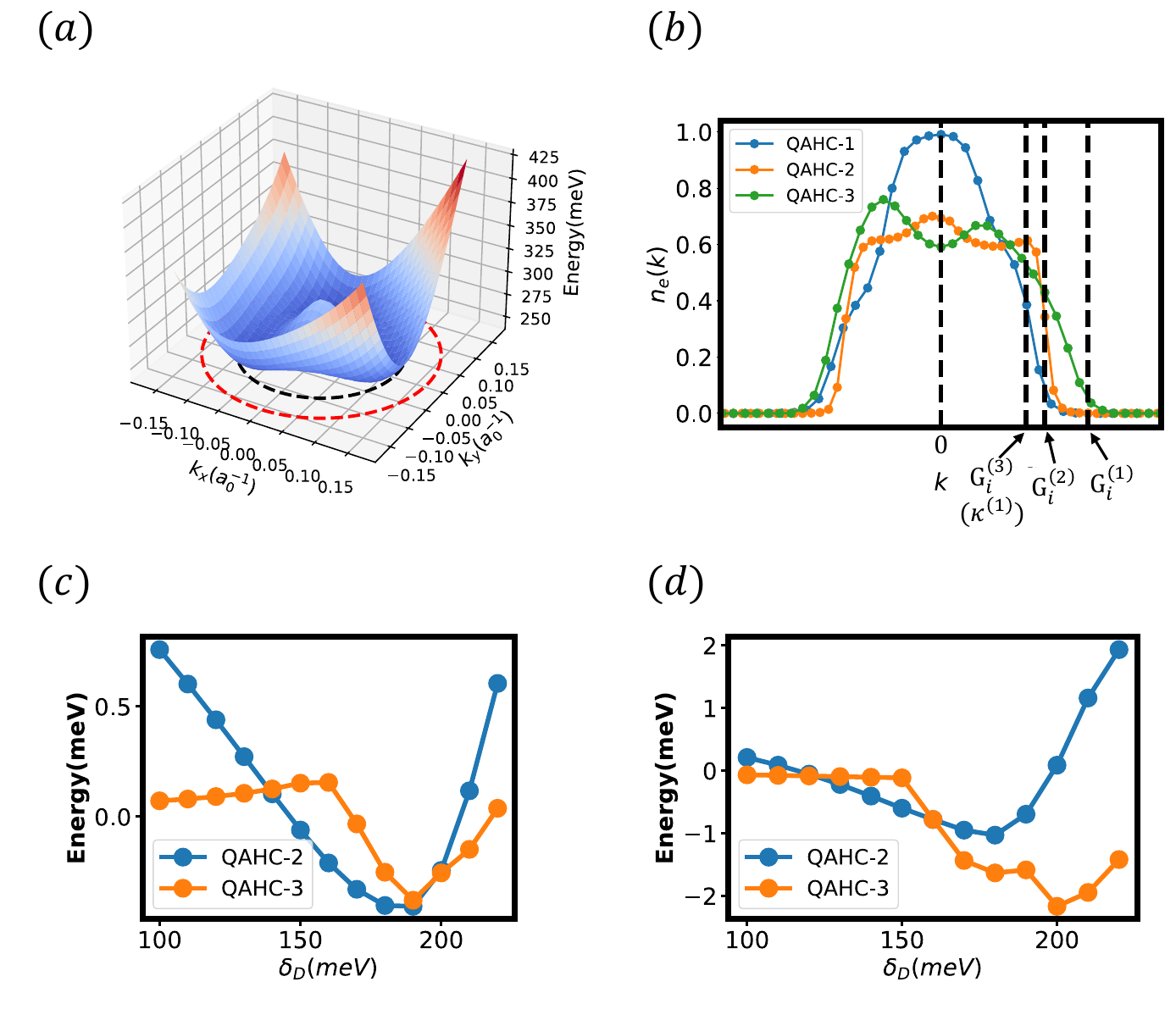}
    \caption{The results for $\theta=0.89^\circ$ ($a_M=10.7$~nm). (a) The conduction band of pentalayer graphene in the moir\'eless limit at $\delta_D=160$~meV. The black and red dashed circle represent $|\mathbf{k}|=|\mathbf{G}^{(2)}_i|$ and $|\mathbf{k}|=|\mathbf{G}^{(1)}_i|$ respectively. (b) The linecut of the momentum distribution $n_e(\mathbf{k})$ at $\delta_D=160$~meV. The blue, orange, green line represents QAHC-$1$, QAHC-$2$, QAHC-$3$ respectively. (c) The total energy and (d) the kinetic energy per moir\'e unit cell of QAHC-$z$ as a function of $\delta_D$. The blue line represents $E_\mathrm{QAHC2}-E_\mathrm{QAHC1}$, the orange line represents $E_\mathrm{QAHC3}-E_\mathrm{QAHC1}$.}
   % (a) The ratio $\frac{\sqrt{2}|\mathbf{k_\mathrm{min}}|}{|\mathbf{G_i}|}$. Here $\mathbf{k_\mathrm{min}}$ is the band bottom of pentalayer graphene.
    \label{fig:band_minimum}
\end{figure}

\textbf{Effect of the kinetic energy} The emergence of enlarged unit cell at $\nu=1$ is a surprise. Here we provide some intuition. The preference for QAHC-$z$ with $z > 1$ arises from the kinetic energy. The band dispersion exhibits a Mexican-hat structure in the original BZ of graphene, with the band minimum forming a ring, as illustrated in Fig.~\ref{fig:band_minimum}(a). Similar results have also been reported in bilayer graphene ~\cite{joy2023wigner}. The kinetic energy of the conduction electrons can be reduced by shifting the momentum distribution, $n_e(\mathbf{k}) = \sum_l \langle c^\dagger_l(\mathbf{k}) c_l(\mathbf{k}) \rangle$, at $\mathbf{k}$ to $\mathbf{k} + \mathbf{G}_i^{(z)}$ in QAHC-$z$.  Here $c_l(\mathbf{k})$ represents the electron operator in the graphene BZ for layer $l$. In the region where QAHC-2 and QAHC-3 are favored, $\mathbf{G}_i^{(z)}$ is located around the ring of the dispersion minimum. Therefore, $n_e(\mathbf k)$ at the $\Gamma$ point of the MBZ is transferred to around the dispersion minimum,  reducing the total kinetic energy. As shown in Fig.~\ref{fig:band_minimum}(b), the momentum distribution $n_e(\mathbf{k})$ at the $\Gamma$ point is close to 1 for QAHC-1, but is considerably reduced for QAHC-2 and QAHC-3. To compare the effects of kinetic energy on total energy, we plot the line cuts of the total energy and kinetic energy of QAHC-$z$ in Fig.~\ref{fig:band_minimum}(c)(d). In the region where QAHC-2 and QAHC-3 are energetically preferred compared to QAHC-1, the kinetic energy is significantly lower for QAHC-2 and QAHC-3 compared to QAHC-1, which compensates a higher interaction energy. We conclude that the QAHC-2 and QAHC-3 phases are unique to the Mexican-hat dispersion. We expect them to be more favored for larger $\epsilon$ with reduced interaction. 

\textbf{Competition between FQAH and IQAH at $\nu=2/3$}
Next we turn to the fractional filling $\nu=2/3$. Previous theories identify an FCI candidate through ED calculation projected onto the lowest band of the QAHC-1 insulator at $\nu=1$~\cite{dong2023theory,zhou2023fractional,dong2023anomalous}. However, the calculation is performed only within a subspace of the full Hilbert space and is better viewed as a variational calculation. Due to the variational nature, the method is biased towards the FCI state and ignore other candidates such as integer QAHCs with a different crystal period. The recent experiments have observed close competition between integer and fractional QAH states at fractional filling~\cite{lu2024extended,waters2024interplay}. To our best knowledge, there is no feasible framework to include all possible crystals in an unbiased way. Note that simply including more Hartree Fock bands from the $\nu=1$ QAHC-1 ansatz  does not really help much, as the multiband calculation is still biased towards the $a_{\text{crystal}}=a_M$ crystal and ignores the possibilities of other crystals with different periods.  Here we still follow the spirit of variational calculations and compare energies of different possible ansatz. 

We target the following ansatzes: (I) an FCI state at fractional filling of the lowest Chern band of the QAHC-1 crystal at $\nu=1$. We obtain this state through an ED calculation. We have improved the procedure to get a self consistent crystal potential directly at the filling $\nu=2/3$~\cite{PhysRevB.110.115146,zhou2023fractional} (see the Supplemental Material).  (II) We obtain various QAHC-$z$ states with integer QAH at $\nu=2/3$ through HF calculations. Now the crystal period is $a_{\text{crystal}}=\sqrt{z/\nu} a_M$ and we still have $z$ bands fully occupied.

\begin{figure}[tbp]
 \includegraphics[width=0.9\linewidth]{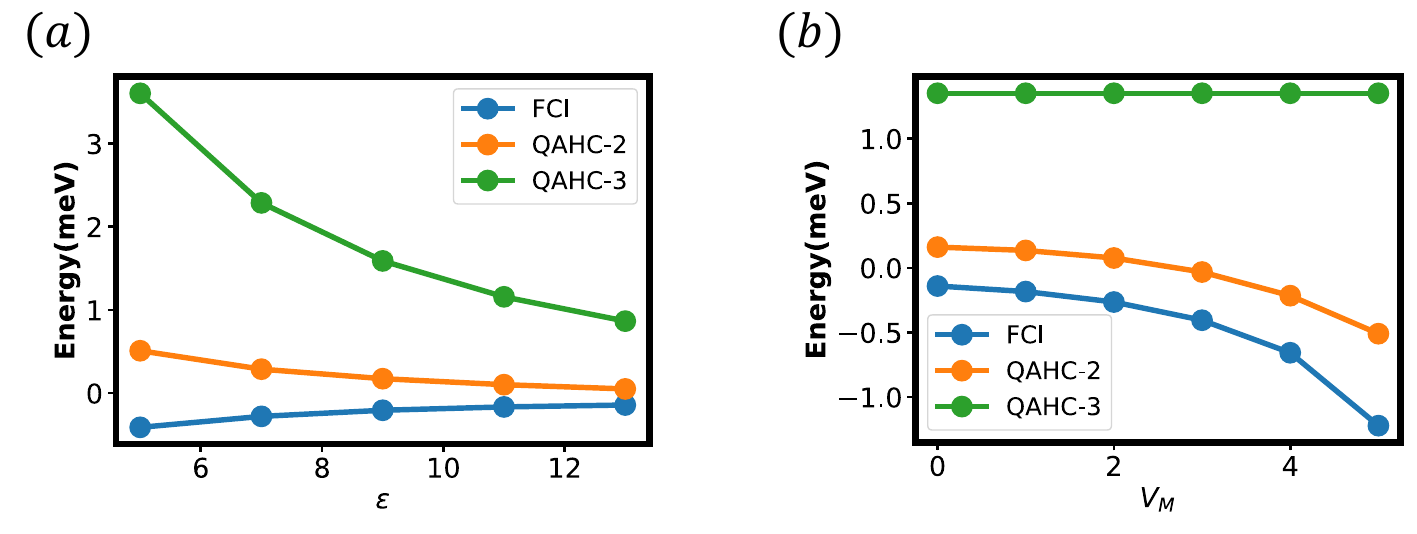}
    \caption{HF and ED calculation at $\delta_D=130$~meV, $\theta=0.89^\circ$ ($a_M=10.7$~nm). (a) The energy per moir\'e unit cell of FCI, QAHC-$2$ and QAHC-$3$ at $\nu=2/3$ as a function of dielectric constant $\epsilon$.  The blue line represents $E_\mathrm{FCI}-E_\mathrm{QAHC1}$, the orange line represents $E_\mathrm{QAHC2}-E_\mathrm{QAHC1}$, the green line represents $E_\mathrm{QAHC3}-E_\mathrm{QAHC1}$. (b) For $\epsilon=10$, dependence of the energy per moir\'e unit cell of FCI, QAHC-$1$, QAHC-$2$ and QAHC-$3$ on moir\'e potential factor $V_M$ at $\nu=2/3$.  }
    \label{fig:change_Vm_epsilon}
\end{figure}

We find the following competing states: FCI, QAHC-1 and QAHC-2. As shown in Fig.~\ref{fig:change_Vm_epsilon}(a), we find that the FCI state is at slightly lower energy than the QAHC-1 state. But when increasing the dielectric constant  $\epsilon$, the integer QAHC-1 and QAHC-2 states become more favored, though still at higher energy than the FCI within our approximation.  Among the integer quantum Hall states, QAHC-2 and QAHC-1 are competing with very close energy. At small $V_M$, QAHC-1 has a lower energy for the $(a_M,D)$ parameter we used.  But with a larger $V_M$, the energy of QAHC-2 becomes lower than that of QAHC-1, due to its commensurability with the moir\'e potential. We note that QAHC-2 is in the same symmetry class as a hole crystal, which can be understood as a trivial charge density wave of holes doped into the QAHC-1 state at $\nu=1$. However, as we discussed in the previous section on $\nu=1$, the QAHC-2 phase we find does not really rely on a parent QAH phase at a larger doping. Instead it may be viewed as a new type of quantum anomalous Hall crystal, which is favored by the moir\'e potential at $\nu=2/3$. 

In summary, we find close competition between FCI and QAHC-1 and QAHC-2 at $\nu=2/3$. Within our approximations, FCI appears to always have lower energy than the integer QAH states. However, the comparison may not be fair as the QHAC-1 and QAHC-2 ansatz are obtained from Hartree Fock calculations. The only confident statement we can make is that the energies of these three states are close. Given the large tunability of the system, all of them might be realized in different parameter regions.  We note that the recent experiments~\cite{lu2024extended,waters2024interplay} observed an integer QAH insulator at $\nu=2/3$. Based on our calculation, we propose either QAHC-1 or QAHC-2 as a possible candidate, which can be only distinguished by imaging the density profile.  

% \yhz{we can delete this paragraph?} In the previous study \cite{dong2023theory,zhou2023fractional,dong2023anomalous}, the moir\'e potential has been considered to be a perturbative potential to pin the interaction-driven QAHC-1. For QAHC-$z$ with $z>1$, the moir\'e potential has a more significant effect on crystals that are commensurate with the moir\'e period. Since the DFT calculation may underestimate the strength of the moiré potential, the pinning effect of the moir\'e potential could be stronger in the actual physical system. Therefore, our calculation may underestimate the energies of crystals that are commensurate with the moir\'e period. To investigate the effect of moir\'e potential more carefully, we increase it manually by adjusting $V_M$ to values larger than $1$ and calculate the energy at $\nu=2/3$. Under the first order perturbation theory, the energies of QAHC-$1$ and QAHC-$3$ remain unchanged due to incommensurability, while the energies of QAHC-$2$ and FQAH state decrease due to the pinning effect, as shown in Fig.~\ref{fig:change_Vm_epsilon}(b). Due to the sensitivity to moir\'e potential, the competition among different QAHC-$z$ and FQAH states remains future experiments to explore. 

\textbf{Summary} In conclusion, we propose a new class of quantum anomalous Hall crystals (QAHCs) with larger crystal periods than the familiar Wigner crystal. Through Hartree Fock calculations, we  demonstrate the possibility of QAHCs with doubled or tripled unit cell even at integer filling $\nu=1$ in pentalayer graphene aligned with hBN.   We also show that the FCI state and two QAHCs are in close competitions at the fractional filling $\nu=2/3$, suggesting a rich phase diagram.  In our study we also notice the important role of the moir\'e potential $V_M$. A larger $V_M$ can greatly change the phase diagram and even favors $C=2$ Chern band at $\nu=1$. We hope this observation motivates more experimental efforts in controlling and enhancing the moir\'e potential, for example through imprinting a superlattice potential from Coulomb interaction of another control layer~\cite{zhou2023fractional}.

\textbf{Acknowledgement} BZ and YHZ thank Dacen Waters and Matthew Yankowitz for previous collaboration on related experimental work. This work was supported by the National Science Foundation under Grant No. DMR-2237031.

\bibliographystyle{apsrev4-1}
\bibliography{paper}

\appendix

\widetext \vspace{0.5cm}

\section{Projection of moir\'e potential}
To compare the energy with different $V_M$, we need a consistent energy background. In the charge neutrality scheme, the vacuum is defined by fully occupying all valence bands of the bare Hamiltonian $H_K=H_0+V_MH_M$. However, the vacuum energy varies with different $V_M$. There are two ways to address this: (1) Moir\'eless charge neutrality scheme: This approach defines the vacuum by occupying all valence bands in the moir\'eless limit, as described in Ref.~\cite{PhysRevB.110.115146}. However, the leading order contribution from the moir\'e potential diverges, as shown in Fig.~2(b) of Ref.~\cite{PhysRevB.110.115146}. Therefore, we consider the second way, which is using moir\'e charge neutrality scheme with projection: We still use the moir\'e charge neutrality scheme, but project the moir\'e potential onto the conduction bands. This ensures that the vacuum energy remains consistent for different $V_M$.

The original moir\'e potential is given by:
\begin{eqnarray}
    H_M=\sum_\mathbf{k}\sum_{j=1}^6 \psi^\dagger_1(\mathbf{k})H_M(\mathbf{G_j})\psi_1(\mathbf{k+G_j}),
\end{eqnarray}
where $H_M(\mathbf{G_j})$ is defined in Eq.~(6) of the main text, and $\psi_z(\mathbf{k})=(f_{z;A}(\mathbf{k}),f_{z;B}(\mathbf{k}))^T$ represents the spin-valley polarized electron operator for the two sublattice in layer $z$. Thre is only $\psi_1(\mathbf{k})$ in $H_M$ because the moir\'e potential only acts on the first layer aligned with the hBN.

To obtain the projection operator, we diagonalize the moir\'eless pentalayer graphene Hamiltonian $H_0$, defined in Eq.~(2) of the main text for each $\mathbf{k}$. This yields the matrix form of the projection operator for conduction electrons at each $\mathbf{k}$, denoted as $P_c(\mathbf{k})$. The projected moiré potential is then:
\begin{eqnarray}
    \tilde{H}_M=\sum_\mathbf{k}\sum_{j=1}^6 \psi^\dagger_1(\mathbf{k})P_c(\mathbf{k})H_M(\mathbf{G_j})P_c(\mathbf{k+G_j})\psi_1(\mathbf{k+G_j}).
\end{eqnarray}
\section{Hartree Fock calculation}\label{sec:self_consistent_ED}
We perform Hartree Fock calculation by keeping the first $N_{b,c}$ conduction bands and the first $N_{b,v}$ valence bands in the MBZ. The interaction term can be written as:
\begin{eqnarray}
    H_V=\frac{1}{2A}\sum_\mathbf{q}\sum_{l,l^\prime}V_{l,l^\prime}(\mathbf{q}):\rho_l(\mathbf{q})\rho_{l^\prime}(-\mathbf{q}):,
\end{eqnarray}
where $\rho_l(\mathbf{q})=\sum_{\mathbf{k},m,m^\prime}c^\dagger_m(\mathbf{k+q})\Lambda^l_{m,m^\prime}(\mathbf{k},\mathbf{q})c_{m^\prime}(\mathbf{k})$ represents the density at layer $l$. Here $\Lambda^l_{m,m^\prime}(\mathbf{k},\mathbf{q})$ is defined as $\langle u_m(\mathbf{k+q})|P_l|u_{m^\prime}(\mathbf{k})\rangle$, $P_l$ is the projection operator to layer $l$, $m$ and $m^\prime$ are band indices. For $m=0,1,...,N_{b,v}-1$, the indices correspond to the valence bands. For $m=N_{b,v},N_{b,v}+1,...,N_{b,v}+N_{b,c}-1$, the indices correspond to the conduction bands. The interaction is given by:
\begin{eqnarray}
    V_{l,l^\prime}(\mathbf{q})=\frac{e^2e^{-q|l-l^\prime|d_\mathrm{layer}}\tanh{(|\mathbf{q}|\lambda)}}{2\epsilon\epsilon_0|\mathbf{q}|},
\end{eqnarray}
where $d_\mathrm{layer}$ is the distance between adjacent layers, $\lambda$ is the screening length. We choose $d_\mathrm{layer}=0.34$~nm and $\lambda=30$~nm in our calculation. The interaction can be decoupled into Hartree and Fock terms, leading to a mean field Hamiltonian:
\begin{eqnarray}
    H_\mathrm{HF}=&\sum_\mathbf{k_1,k_2}\sum_{m,m^\prime,n,n^\prime}2\left(V_{m,m^\prime,n,n^\prime}(\mathbf{k_1,k_2,0})-V_{n,m^\prime,m,n^\prime}(\mathbf{k_1,k_2,k_2-k_1})\right)&\\
    &\left(\langle c^\dagger_m(\mathbf{k_1})c_{m^\prime}(\mathbf{k_1})\rangle-\delta_{m,m^\prime,\mathrm{valence}}\right)c^\dagger_n(\mathbf{k_2})c_{n^\prime}(\mathbf{k_2}),&
    \label{Eq:self_consistent}
\end{eqnarray}
where $m,m^\prime,n,n^\prime$ are band indices. $\delta_{m,m^\prime,\mathrm{valence}}=1$ while $m=m^\prime$ and $m$ is valence band, otherwise it is $0$. We include this term because we use the charge neutrality scheme, in which the reference state is all of the valence bands are occupied. The expectation value $\langle c^\dagger_m(\mathbf{k_1})c_{m^\prime}(\mathbf{k_1})\rangle$ is calculated by using the free fermion mean field state given by $H_K+H_\mathrm{HF}$, with all valence bands and $z$ conduction bands occupied. Here $z$ corresponds to the QAHC-$z$ ansatz. The interaction vertex is calculated as:
\begin{eqnarray}
    V_{m,m^\prime,n,n^\prime}(\mathbf{k_1},\mathbf{k_2},\mathbf{q})=\frac{1}{2A}\sum_{l,l^\prime}V_{l,l^\prime}(\mathbf{q})\Lambda^l_{m,m^\prime}(\mathbf{k_1},\mathbf{q})\Lambda^{l^\prime}_{n,n^\prime}(\mathbf{k_2},\mathbf{-q}).
\end{eqnarray}
The mean field energy is calculated as:
\begin{eqnarray}
    E_\mathrm{MF}=&&\sum_{\mathbf{k},n}\epsilon_n(\mathbf{k})\left(\langle c^\dagger_n(\mathbf{k}) c_n(\mathbf{k})\rangle-\delta_{n,\mathrm{valence}}\right)\\
   &+&\sum_{\mathbf{k_1},\mathbf{k_2}}\sum_{m,m^\prime,n,n^\prime}\left(V_{m,m^\prime,n,n^\prime}(\mathbf{k_1,k_2,0})-V_{n,m^\prime,m,n^\prime}(\mathbf{k_1,k_2,k_2-k_1})\right)\\
   &&\left(\langle c^\dagger_m(\mathbf{k_1})c_{m^\prime}(\mathbf{k_1})\rangle-\delta_{m,m^\prime,\mathrm{valence}}\right)\left(\langle c^\dagger_n(\mathbf{k_2})c_{n^\prime}(\mathbf{k_2})\rangle-\delta_{n,n^\prime,\mathrm{valence}}\right),
\end{eqnarray}
in which $\delta_{n,\mathrm{valence}}=1$ while $n$ is valence band, otherwise it is $0$. $\epsilon_n(\mathbf{k})$ is the dispersion of the $n^\mathrm{th}$ band of $H_K$.
In our HF calculation, we try 40 number of randomized initial ansatz and choose the one with the lowest energy.
\section{ED calculation}
We perform ED calculation by projecting the Coulomb interaction into the lowest conduction HF band, which is the $\left(N_{b,v}+1\right)^\mathrm{th}$ lowest HF band. The total Hamiltonian is expressed as:
\begin{equation}
    H=\sum_\mathbf{k} \epsilon(\mathbf{k})\tilde{c}_{N_{b,v}}^\dagger(\mathbf{k})\tilde{c}_{N_{b,v}}(\mathbf{k})+\sum_{\mathbf{k_1},\mathbf{k_2},\mathbf{q}}V(\mathbf{k_1},\mathbf{k_2},\mathbf{q})\tilde{c}_{N_{b,v}}^\dagger(\mathbf{k_1+q})\tilde{c}_{N_{b,v}}^\dagger(\mathbf{k_2-q})\tilde{c}_{N_{b,v}}(\mathbf{k_2})\tilde{c}_{N_{b,v}}(\mathbf{k_1}),
\end{equation}
where $\tilde{c}_{N_{b,v}}^\dagger(\mathbf{k})=\sum_n U_{N_{b,v}n}(\mathbf{k})c_n^\dagger(\mathbf{k})$ represents the electron operator of the lowest conduction HF band, with $c_n^\dagger(\mathbf{k})$ being the electron operator in bare Hamiltonian $H_K$, and $n$ indexing the bands. The dispersion $\epsilon(\mathbf{k})$ is obtained by projecting the dispersion of the kinetic term $H_K=H_0+V_M H_M$ onto the lowest conduction HF band:
\begin{equation}
    \epsilon(\mathbf{k})=\sum_n |U_{N_{b,v}n}(\mathbf{k})|^2\epsilon_n(\mathbf{k}),
\end{equation}
where $\epsilon_n(\mathbf{k})$ is the dispersion of the $n^\mathrm{th}$ band of $H_K$. The interaction vertex takes the form:
\begin{equation}\label{eqVertex}
    V(\mathbf{k_1},\mathbf{k_2},\mathbf{q})=\frac{1}{2A}\sum_{l,l^\prime} V_{l,l^\prime}(\mathbf{q})\Lambda^l_\mathrm{HF}(\mathbf{k_1},\mathbf{q})\Lambda^{l^\prime}_\mathrm{HF}(\mathbf{k_2},\mathbf{-q}),
\end{equation}
where $\Lambda^l_\mathrm{HF}(\mathbf{k_1},\mathbf{q})=\langle u_{\mathrm{HF},N_{b.v}}(\mathbf{k_1+q})|P_l|u_{\mathrm{HF},N_{b,v}}(\mathbf{k_1})\rangle$ is the form factor of the lowest conduction HF Bloch wavefunction on layer $l$.

Applying the particle-hole transformation $\tilde{c}_{N_{b,v}}(\mathbf{k})\rightarrow \tilde{h}_{N_{b,v}}^\dagger(\mathbf{k})$, we obtain the total Hamiltonian including the kinetic and interaction term:
\begin{equation}\label{eqED}
\begin{split}
    H=&-\sum_\mathbf{k}\left( \epsilon(\mathbf{k})+2\sum_{\mathbf{k^\prime}}\left(V(\mathbf{k},\mathbf{k^\prime},\mathbf{0})-V(\mathbf{k},\mathbf{k^\prime},\mathbf{k^\prime-k})\right)\right)\tilde{h}_{N_{b,v}}^\dagger(\mathbf{k})\tilde{h}_{N_{b,v}}(\mathbf{k})\\
    &+\sum_{\mathbf{k_1},\mathbf{k_2},\mathbf{q}}V(\mathbf{k_1},\mathbf{k_2},\mathbf{q})\tilde{h}_{N_{b,v}}^\dagger(\mathbf{k_1+q})\tilde{h}_{N_{b,v}}^\dagger(\mathbf{k_2-q})\tilde{h}_{N_{b,v}}(\mathbf{k_2})\tilde{h}_{N_{b,v}}(\mathbf{k_1})+E_0.
\end{split}
\end{equation}
Here $\epsilon(\mathbf{k})-\sum_{\mathbf{k^\prime}}\left(V(\mathbf{k},\mathbf{k^\prime},\mathbf{0})-V(\mathbf{k},\mathbf{k^\prime},\mathbf{k^\prime-k})\right)$ equals to the HF dispersion. $E_0$ is the energy of the $\nu=1$ insulating state:
\begin{equation}\label{eqE0}
    E_0=\sum_\mathbf{k} \left(\epsilon(\mathbf{k}) + \sum_{\mathbf{k^\prime}}\left(V(\mathbf{k},\mathbf{k^\prime},0)-V(\mathbf{k},\mathbf{k^\prime},\mathbf{k^\prime-k})\right)\right).
\end{equation}
Therefore, we are using the hole picture relative to the $\nu=1$ state, applying the dispersion and the form factor of the HF band to perform the ED calculation.

In the ED calculation, we define the many body Hamiltonian in the hole picture relative to the $\nu=1$ spin-valley polarized Chern insulator. We stress that performing the calculation in the electron picture will lead to double-counting issues. In our main text, $\nu=2/3$ filling is treated as $1/3$ filling in terms of holes relative to the $\nu=1$ Chern insulator. Since the $\nu=1$ parent state is spin-valley polarized, the hole creation operator of the other spin or valley species is meaningless, allowing us to access only spin-valley polarized FCI states within this framework. 

\section{Self consistent ED calculation at fractional filling beyond the rigid band approximation}
We go beyond the rigid band approximation and perform a self consistent calculation at $\nu=2/3$ directly. The procedure is similar to the standard self consistent HF calculation in Eq.~\ref{Eq:self_consistent}, but in this case, the expectation value $\langle c^\dagger_m(\mathbf{k_1}) c_{m^\prime}(\mathbf{k_1})\rangle$ is directly obtained from the many body FCI state at $\nu=2/3$, instead of a free fermion mean field state. During each iteration, we get bands from the mean field Hamiltonian $H_K+H_\mathrm{HF}$, where the creation operator $\tilde{c}^\dagger_m(\mathbf{k})$ corresponds to the $m=0,1,...$ bands.

An FCI state at $\nu=2/3$ can be constructed for the band generated by $\tilde c_{N_{b,v}}(\mathbf k)$. In principle, we can use ED to get the FCI state and calculate the expectation value,  which enters the self consistent equation in Eq.~\ref{Eq:self_consistent} to generate a new $H_\mathrm{MF}$. However, we simplify the process by noting that the FCI state has an almost uniform momentum distribution. Thus, we approximately have:
\begin{equation}
    \langle\tilde{c}^\dagger_m(\mathbf{k_1})\tilde{c}_{m^\prime}(\mathbf{k_1})\rangle\approx \nu\delta_{m,N_{b,v}}\delta_{m^\prime,N_{b,v}},
\end{equation}
where $\tilde{c}^\dagger_m(\mathbf{k})=\sum_n U_{mn}(\mathbf{k})c^\dagger_n(\mathbf{k})$ is the electron operator of the $m^\mathrm{th}$ mean field band and $c^\dagger_n(\mathbf{k})$ is the electron operator in the band of the bare $H_K$, with $n$ is the bare band index. The expectation value $\langle c^\dagger_{m}(\mathbf{k_1})c_{m^\prime}(\mathbf{k_1})\rangle$ which enters Eq.~\ref{Eq:self_consistent} can now be calculated as:
\begin{equation}
    \langle c^\dagger_m (\mathbf{k_1}) c_{m^\prime}(\mathbf{k_1})\rangle\approx\nu U^*_{N_{b,v}m}(\mathbf{k_1}) U_{N_{b,v}m^\prime}(\mathbf{k_1}).
\end{equation}
\section{Integer QAHC-$z$ at $\nu=2/3$}
Besides the FQAH states at $\nu=2/3$, we also target the integer QAHC-$z$ as variational ansatzes. As discussed in the main text, for $z/\nu=m_1^2+m_2^2+m_1m_2$ with $m_1,m_2\in\mathbb{Z}$, the moir\'e potential is commensurate with the crystal period. We consider $z=1,2,3$, only QAHC-$2$ is commensurate with the moir\'e potential. In this case, $a_\mathrm{crystal}=\sqrt{z/\nu}a_M=\sqrt{3}a_M$. Suppose the reciprocal vector of the QAHC-$2$ crystal at $\nu=2/3$ is $\mathbf{g}_j$, $j=1,2,3,4,5,6$. The relation between $\mathbf{G}_j$ and $\mathbf{g}_j$ is:
\begin{eqnarray}
    \mathbf{G}_1&=&\mathbf{g}_1-2\mathbf{g}_2,\\
    \mathbf{G}_2&=&2\mathbf{g}_1-\mathbf{g}_2,\\
    \mathbf{G}_3&=&\mathbf{g}_1+\mathbf{g}_2,\\
    \mathbf{G}_4&=&-\mathbf{g}_1+2\mathbf{g}_2,\\
    \mathbf{G}_5&=&-2\mathbf{g}_1+\mathbf{g}_2,\\
    \mathbf{G}_6&=&-\mathbf{g}_1-2\mathbf{g}_2.
\end{eqnarray}
Then we can write the moir\'e potential in the basis of $\mathbf{g}_j$ and perform the calculations.
\section{Details of QAHC-z energy}
\begin{figure}[tbp]
\centering
    \includegraphics[width=0.8\linewidth]{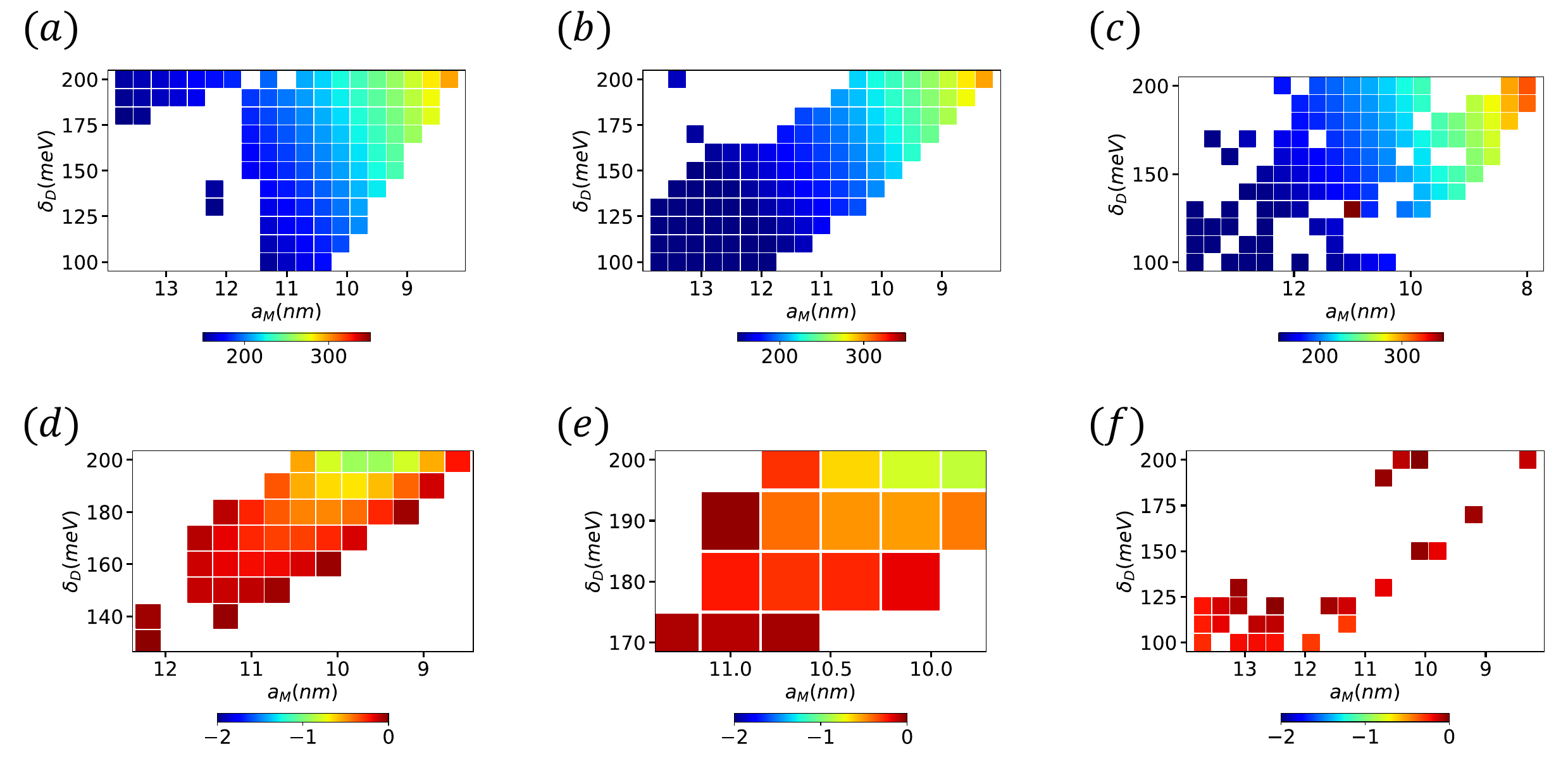}
    \caption{The energy per moir\'e unit cell for (a) QAHC-$1$, (b) QAHC-$2$ and (c) QAHC-$3$. The energy difference per moir\'e unit cell for (d) $E_\mathrm{QAHC2}-E_\mathrm{QAHC1}$, (e) $E_\mathrm{QAHC3}-E_\mathrm{QAHC1}$ and (f) $E_\mathrm{QAHC3}-E_\mathrm{QAHC2}$.}
    \label{fig:AHCn_energy}
\end{figure}

We present the energy per moir\'e unit cell for QAHC-$z$ with $z=1,2,3$ in Fig.~\ref{fig:AHCn_energy}(a), (b), (c) respectively. The blank regions correspond to the metal phase or the trivial Wigner crystal phase. The calculations were conducted with $\epsilon=10$. For QAHC-$1$ and QAHC-$2$, we keep $N_{b,v}=0$ valence bands and $N_{b,c}=7$ conduction bands, using a $12\times12$ grid in the crystal Brillouin zone for the HF calculations. For QAHC-$3$, we keep $N_{b,v}=0$ valence bands and $N_{b,c}=9$ conduction bands, using a $9\times9$ grid in the crystal Brillouin zone for the HF calculations. We also show the energy difference of different QAHC states in Fig.~\ref{fig:AHCn_energy}(d)-(f), in which we only keep the negative value.

\section{Demonstration of convergence in HF calculation}
\begin{figure}[tbp]
 \centering
    \includegraphics[width=0.65\linewidth]{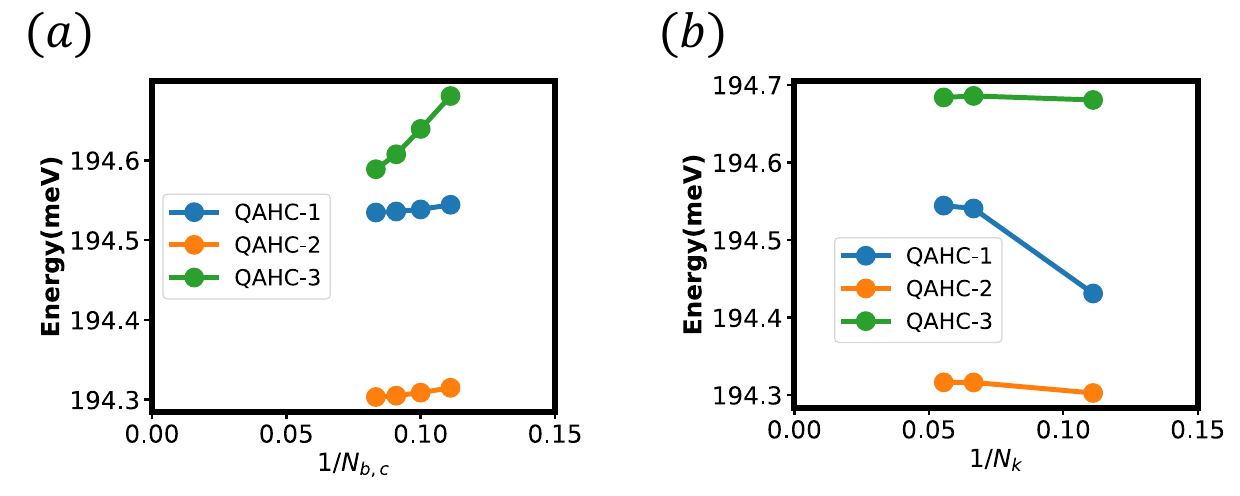}
    \caption{(a) The energy per moir\'e unit cell of QAHC-$z$ varies with $1/N_{b,c}$. The energy difference converges to a positive value as $N_{b,c}\rightarrow\infty$. For $z=1,2,3$, we choose $N_k=18,12,9$ respectively. (b) The energy per moir\'e unit cell for QHAC-$z$ with $z=1,2,3$ is plotted as a function of $1/N_k$. We choose $N_{b,c}=9$ for $z=1,2,3$. }
    \label{fig:finite_size}
\end{figure}

In the main text, we set $N_{b,v} = 0$ and $N_{b,c} = 7, 7, 9$ for QAHC-1, QAHC-2, and QAHC-3 respectively to perform the Hartree Fock calculation. The calculations are performed using an $N_k \times N_k$ grid in the crystal BZ, with $N_k$ set to 12, 12, and 9 for $z = 1, 2, 3$, respectively. To demonstrate the convergence of our calculation, we select specific parameters with $\delta_D = 160$ meV and $\theta = 0.89^\circ$ ($a_M = 10.7$ nm), and then increase $N_{b,c}$ and $N_k$ to perform the calculation, as shown in Fig.~\ref{fig:finite_size}(a)(b) respectively. In the limit $N_{b,c}\rightarrow\infty$ and $N_k\rightarrow\infty$, the QAHC-$2$ state converges to have the lowest energy.

\section{Phase diagram with another parameter}
In other reference~\cite{dong2023anomalous}, people use a different set of parameters. The parameters they used are $(\gamma_0,\gamma_1,\gamma_2,\gamma_3,\gamma_4)=(-3100,380,-21,290,141)$~meV. Compared to Eq.~3 in the main text, they assume the potential difference is uniform in the system and ignore the term of $u_{A,i}$, $u_{B,i}$. For the moir\'e potential term, it is:
\begin{equation}
    H_M(\mathbf{G}_j) = 
    \begin{pmatrix}
        V_0+V_1e^{\mathrm{i}\psi} & V_1e^{\mathrm{i}(\frac{(3-j)\pi}{3}+\psi)}\\
        V_1e^{\mathrm{i}(\frac{(1+j)\pi}{3}+\psi)} & V_0+V_1e^{\mathrm{i}(-\frac{2\pi}{3}+\psi)} 
    \end{pmatrix},
\end{equation}
where the momentum difference given by $\mathbf{G}_j=\frac{4\pi}{\sqrt{3}L_M}(\cos (\frac{j\pi}{3}-\frac{5\pi}{6}),\sin (\frac{j\pi}{3}-\frac{5\pi}{6}))^T$ for $j=1,3,5$. For $j=2,4,6$, the tunneling is obtained by taking the Hermitian conjugate. The parameters are $(V_0,V_1,\psi)=(28.9~\mathrm{meV},21~\mathrm{meV},-0.29)$. 

\begin{figure}[tbp]
 \centering
    \includegraphics[width=0.8\linewidth]{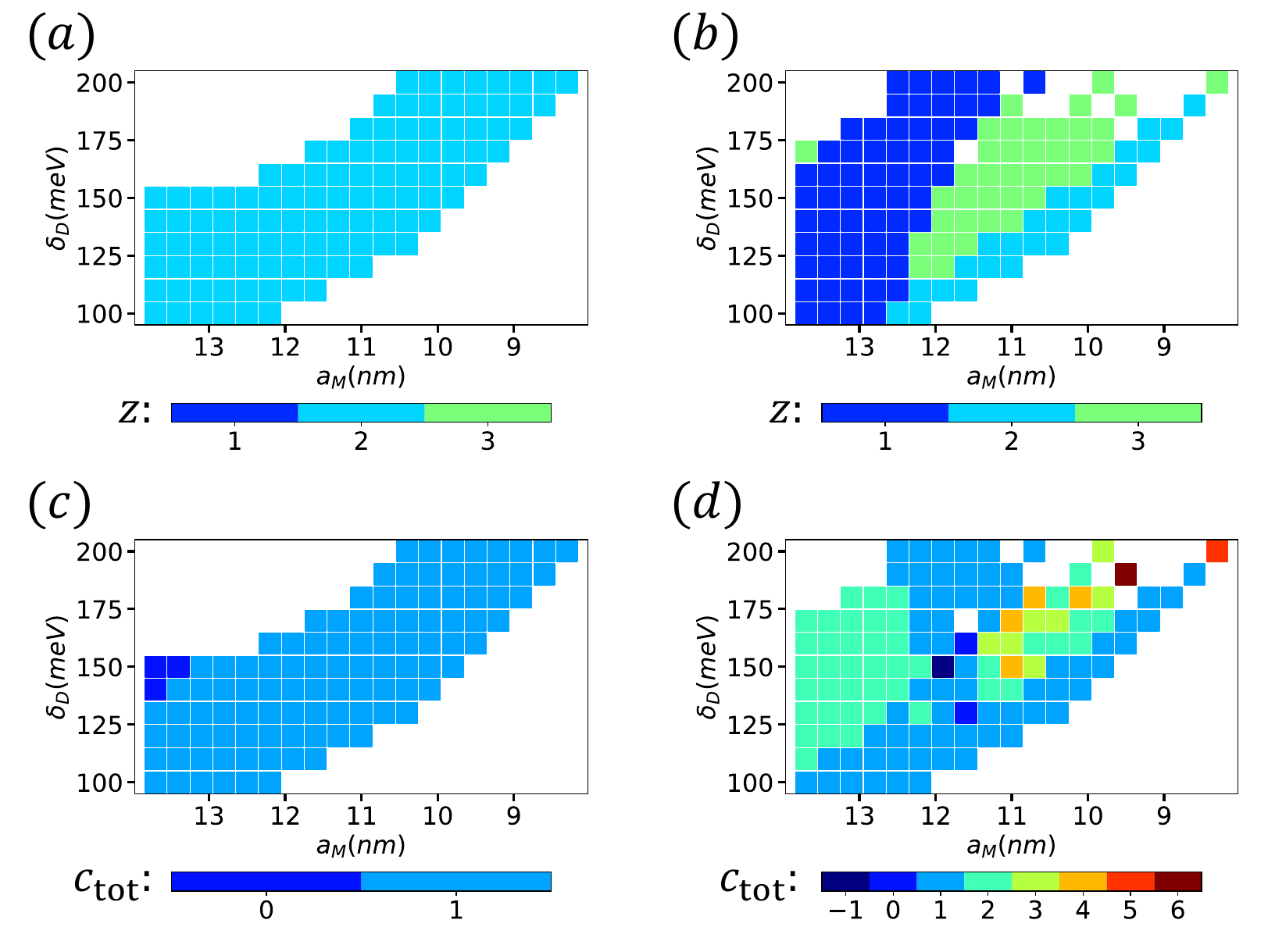}
    \caption{(a) The phase diagram of QAHC as a function of displacement field $D$ and moir\'e period $a_M$. We keep $N_{b,c}$ conduction bands and $N_{b,v}$ valence bands in the calculation. For QAHC-$z$ with $z=1,2,3$, we set $N_{b,c}$ to 7,7 and 9 respectively. $N_{b,v}$ is taken as $0$ for all values of $z$, since the valence bands are separated by the large displacement field. The calculations are performed using an $N_k\times N_k$ grid in the crystal BZ. For $z=1,2,3$, we set $N_k$ to 12, 12 and 9 respectively. (a) Dependence of $z$ on $D$ and $a_M$ for $V_M=1$. (b) Dependence of $z$ on $D$ and $a_M$ for $V_M=2$. (c) Dependence of $C_\mathrm{tot}$ on $D$ and $a_M$ for $V_M=1$. (d) Dependence of $C_\mathrm{tot}$ on $D$ and $a_M$ for $V_M=2$.} 
    \label{fig:phase_diagram_ashvin_parameter}
\end{figure}

We use the new parameter for the free Hamiltonian $H_K$ and retain the same parameters as in the main text, $(\lambda, d_\mathrm{layer}, \epsilon) = (30~\mathrm{nm}, 0.34~\mathrm{nm}, 10)$, for the Coulomb interaction term $H_V$ to perform the Hartree Fock calculation, obtaining the phase diagram at $\nu = 1$ shown in Fig.~\ref{fig:phase_diagram_ashvin_parameter}.

\section{Calculation by single gate screened Coulomb interaction}
In our main text, all HF and ED calculations are performed by using the double gate screened Coulomb interaction. In this section, we repeat the same calculation by using the following single gate screened potential:
\begin{equation}
    V_{ll^\prime}(\mathbf{q})=\frac{e^2(1-e^{-q\lambda})}{2\epsilon\epsilon_0 q}e^{-q|l-l^\prime|d_\mathrm{layer}},
\end{equation}
where $\lambda=30$ nm is the screening length and $d_\mathrm{layer}=0.34$ nm is the distance between the adjacent layers. 

Following the same procedure as in the main text, we perform HF calculations and obtain a phase diagram. The phase diagram is similar to Fig.~1(a) in the main text. Notably, the parameter region where the QAHC-2 state has the lowest energy is larger compared to the case using the double gate screened potential.  We present the phase diagram, along with the corresponding density profile and dispersion of QAHC-$2$ at $\theta=0.89^\circ$ ($a_M=10.7$ nm), $\delta_D=160$ meV in the figure below.

% Following the same procedure as in the main text, we change $\epsilon$ and calculate the energy per particle for the following ansatzes: (I) an FCI state at
% fractional filling of the lowest Chern band of the QAHC-1 crystal at $\nu=1$, obtained through self consistent ED
% calculation as described in Sec.~\ref{sec:self_consistent_ED}. (II) Various QAHC-z states with integer QAH at $\nu=2/3$, obtained via
% HF calculations. The crystal period in this case is $a_\mathrm{crystal}=\sqrt{z/\nu}a_M$ and $z$ bands fully filled. The results are shown in Fig.~ and compared with those obtained using the double ßgate screened potential.
\begin{figure}[tbp]
 \centering
    \includegraphics[width=0.8\linewidth]{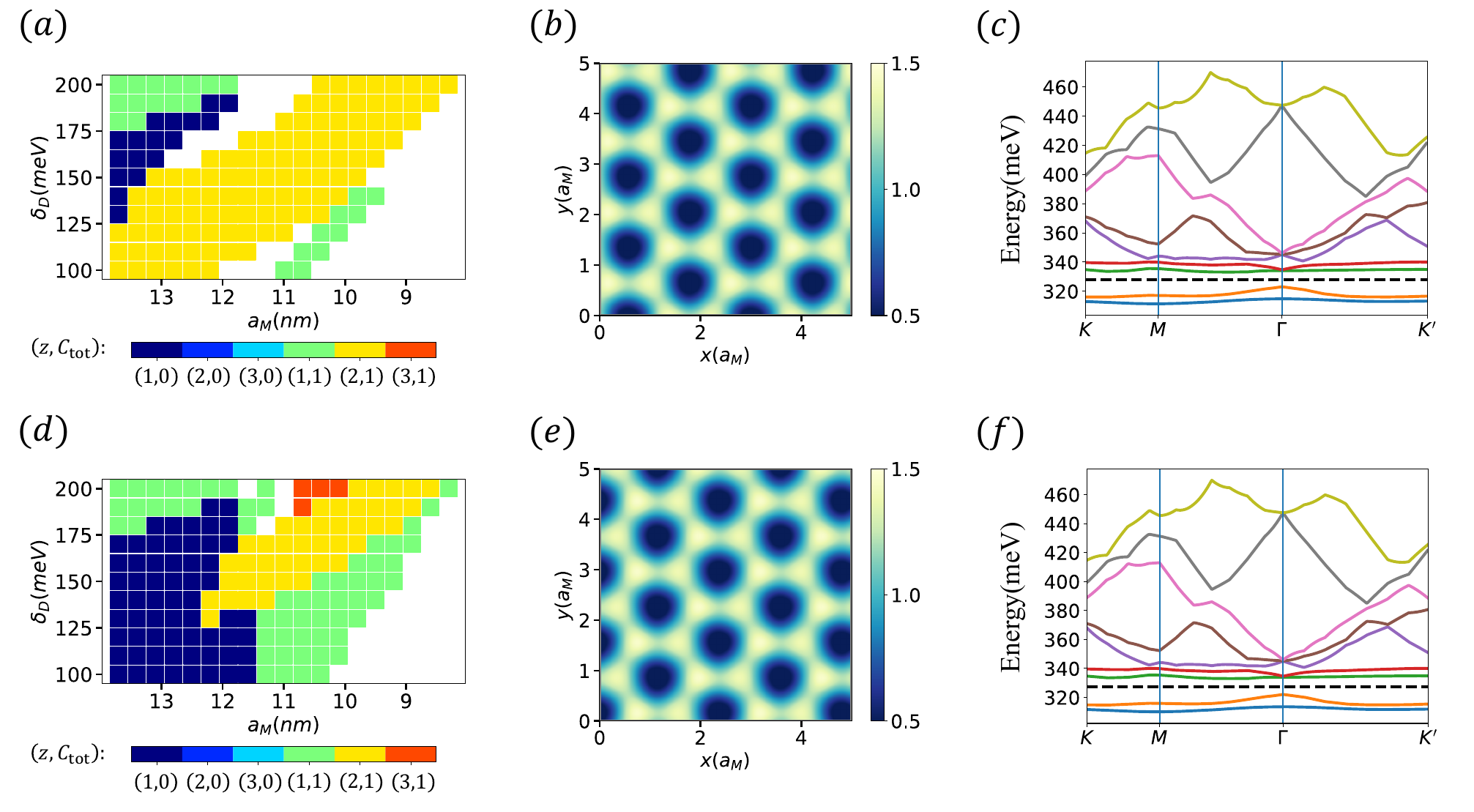}
    \caption{Calculations using the single gate screened potential are shown in (a)–(c), and calculations using the double gate screened potential are shown in (d)–(f). (a) and (d) present the phase diagram of QAHC states as a function of the displacement field $\delta_D$ and moiré period $a_M$, with $V_M=1$. The label $(z, C_\mathrm{tot})$ denotes a QAHC-$z$ state with total Chern number $C_\mathrm{tot}$. (b) and (e) show the density profiles of the QAHC-2 state at $\theta=0.89^\circ$ ($a_M = 10.7$ nm) and $\delta_D = 160$ meV. (c) and (f) display the corresponding HF dispersions for the same parameters as in (b) and (e). In the dispersion plots, the black dashed line indicates the Fermi energy.} 
    \label{fig:single_gate}
\end{figure}

\end{document}